\documentclass[a4paper,11pt]{article}
\usepackage{jheppub} 

\def \N{{\mathcal N}}

\def \Q{{\mathbfcal{Q}}}
\def \q{{\mathbfcal{q}}}
\def \J{{\mathcal J}}
\def \D{{\mathcal D}}
\def \gTr{{{\cal T}\!r}} 
\def \Y{{\mathcal Y}}
\def \gG{{\mathcal G}}
\def \g{{\mathfrak{g}}}

\usepackage{braket}
\def\fnum@figure{\textbf{\figurename\nobreakspace\thefigure}}
\def\fnum@table{\textbf{\tablename\nobreakspace\thetable}}
\usepackage{bm}
\newcommand{\gh}{\mathrm{gh}}

\DeclareMathAlphabet\mathbfcal{OMS}{cmsy}{b}{n}
\usepackage{bbold}

\usepackage{xcolor}
\definecolor{heavypurple}{RGB}{170, 46, 116} 
\definecolor{purple}{RGB}{246, 92, 179} 
\definecolor{orange}{RGB}{255, 138, 54}
\definecolor{yellow}{RGB}{244, 196, 48}
\definecolor{lightblue}{RGB}{79, 139, 190}

\usepackage{cancel}
\usepackage{mathtools}

\usepackage[normalem]{ulem}

\title{\boldmath \Huge Massive gravity from a first-quantized perspective}

\author{Filippo Fecit}
\affiliation{Dipartimento di Fisica e Astronomia ``Augusto Righi", Universit{\`a} di Bologna,\\
via Irnerio 46, I-40126 Bologna, Italy}
\affiliation{INFN, Sezione di Bologna, \\
via Irnerio 46, I-40126 Bologna, Italy}

\emailAdd{filippo.fecit2@unibo.it}

\abstract{In this work, we investigate the BRST quantization of the massive $\N=4$ supersymmetric spinning particle, with a twofold purpose: exploring different approaches to give mass to spinning particle models and formulating a first-quantized theory for linearized massive gravity on both flat and curved spacetime. Our results suggest that achieving the nilpotency of the BRST charge requires an Einstein spacetime with vanishing cosmological constant as the only viable consistent background. In the course of the analysis, we take the $\N=2$ supersymmetric worldline as an exemplificative model, correctly producing the Proca theory on curved spacetime. Our analysis shows that the associated BRST system uniquely selects the minimal coupling to the background curvature.}

\makeatletter
\gdef\@fpheader{}
\makeatother

\begin{document}
\maketitle
\flushbottom

\section{Introduction}
The theory of General Relativity (GR) has been the subject of modification attempts since the early years following Einstein's formulation. Despite the considerable success of GR in the description of fundamental aspects of our reality, several phenomena persist unexplained, prompting the study of new physics. An intriguing modification of gravitational theories suggests that gravity is propagated by a \emph{massive} spin $2$ particle: a massive graviton. 
The question of whether the graviton possesses mass remains an unresolved inquiry to date. This prospect, far from being dismissible on the basis of theoretical and experimental arguments \cite{deRham:2016nuf}, would entail profound consequences, which may account for the enduring history of theoretical explorations dedicated to this possibility. The first theory of massive gravity dates back to 1939 when Fierz and Pauli proposed a relativistic action for a massive spin $2$ particle on flat spacetime which is now known as Fierz-Pauli (FP) theory \cite{Fierz:1939zz, Fierz:1939ix}. While appearing deceptively simple in its formulation, the theory already reveals the roots of several issues that have consistently plagued every effort to improve massive gravity beyond the linear level. Specifically, two primary obstacles arise: the vDVZ discontinuity \cite{vanDam:1970vg, Zakharov:1970cc}, namely the different predictions between linear GR and the FP theory in the massless limit, and the presence of a Boulware-Deser (BD) ghost, an extra degree of freedom which inevitably appears at the non-linear level \cite{Boulware:1972yco}. To this date, only a promising non-linear theory of massive gravity free from the DB ghost instability has been formulated, which is known as dRGT theory \cite{deRham:2010ik, deRham:2010kj, deRham:2011rn}. The interested reader may refer to excellent reviews for a comprehensive exploration of this subject \cite{Hinterbichler:2011tt, deRham:2014zqa, Schmidt-May:2015vnx}.
\\
This paper represents the first attempt to contribute to the ongoing investigation from an alternative perspective by employing a worldline approach \cite{Schubert:2001he}. The primary goal is to achieve a first-quantized description of massive gravity. To this end, we exploit the first-quantized models known as 
$O(\N)$ spinning particles. These are mechanical models with $\N$ local supersymmetry on the worldline that have been shown to describe, on Minkowski, a spin $s=\frac{\N}{2}$ particle in first quantization and constitute an alternative to conventional second-quantized field theories in the study of quantum field theory (QFT) \cite{Berezin:1976eg, Brink:1976uf, Gershun:1979fb, Howe:1988ft, Howe:1989vn, Siegel:1988ru, Bastianelli:2008nm, Corradini:2010ia, Bastianelli:2014lia, Corradini:2015tik}. In recent years, considerable effort has been invested in exploring the consistent coupling of the spinning particles to more general backgrounds beyond flat spacetime. 
It was shown in \cite{Howe:1988ft} that for $\N>2$, worldline supersymmetry transformation rules leave the spinning particle action invariant only if the target spacetime is flat. Subsequent progress was made in \cite{Kuzenko:1995mg} and in \cite{Bastianelli:2008nm}, where it was established how to couple the spinning particle to (A)dS spaces and arbitrary conformally flat spaces respectively. A significant advance has been recently achieved realizing that the technique of BRST quantization -- originally developed in the context of the path integral quantization of Lagrangian gauge theories \cite{Becchi:1975nq} -- provides a way to investigate different backgrounds rather efficiently. An appealing aspect of BRST quantization relies on the automatic generation of the complete spectrum of spacetime fields essential for the Batalin-Vilkoviski (BV) quantization of the corresponding spacetime quantum field theory \cite{Batalin:1981jr, Batalin:1983ggl}. The BRST quantization of the spinning particle models led to a first-quantized description of Yang-Mills theory from the $\N=2$ spinning particle \cite{Dai:2008bh} and Einstein gravity from the $\N=4$ spinning particle \cite{Bonezzi:2018box}. The latter analysis showed quantum consistency in coupling the model to a more general set of backgrounds, namely to Einstein spaces. Furthermore, this formalism has been extended to include the NS-sector of supergravity in \cite{Bonezzi:2020jjq}. \\
In this work, we intend to apply this procedure to reproduce linearized massive gravity (LMG) on both flat and curved spacetime. As a warm-up, we address the case of the massive $\N=2$ spinning particle and its connection to the Proca theory. This preliminary investigation serves a dual purpose: introducing the relevant methodologies and exploring the potential impact of the mass as an obstruction to the BRST algebra. One viable approach to incorporate a mass term is the Scherk-Schwarz mechanism \cite{Scherk:1979zr}, which entails a dimensional reduction of the massless model. Recently, another method has been developed consisting of the introduction of auxiliary oscillators describing the internal degrees of freedom of the spinning particle \cite{Carosi:2021wbi}. In this work, both methods are employed to confer mass to the graviton on a flat background, allowing for an in-depth exploration of the main features of the \emph{massive} $\N=4$ supersymmetric worldline theory. Notably, these two approaches appear to yield different outcomes, with the first method providing the correct BV spectrum and field equations of linearized massive gravity. Conversely, the second approach encounters difficulties already at the initial stage. Subsequently, we attempt to couple the model to a generic curved background, and our findings suggest that the nilpotency of the BRST charge requires the background to be Einstein with cosmological constant set to zero. \\

The paper is organized as follows. In section \ref{sec1} we outline the main features of the $O(\N)$ massless spinning particle models, providing the basis for the upcoming analysis. Section \ref{sec2} provides an overview of the BRST quantization procedure, with emphasis on the first-quantized massless graviton both on flat and curved spacetime. In section \ref{sec3} we discuss how to confer a mass to a spinning particle model. We begin by taking the $\N=2$ spinning particle as an illustrative example to elucidate the role of the mass in the BRST algebra, showing that the model correctly reproduces the Proca theory on a general background. We then proceed to give a mass to the graviton and check whether the theory correctly reproduces that of Fierz-Pauli. Finally, in section \ref{sec4} we attempt to couple the model to a generic curved background, establishing the consistency of the BRST quantization exclusively on a subset of viable spacetimes. Conclusions and possible outlooks are presented in section \ref{conc}. 


\section{$O(\N)$ massless spinning particles} \label{sec1}
The so-called $O(\N)$ spinning particles represent prominent examples where gauge symmetries allow for a manifestly Lorentz covariant formulation. This, in turn, gives rise to a constrained Hamiltonian system with first-class constraints \cite{new-book}. These kinds of constraints, denoted as $C_\alpha$ (with $\alpha$ spanning the number of constraints), satisfy upon quantization a graded algebra of the form\footnote{With the following notation for the graded commutator:
\begin{equation*}
[\cdot,\cdot\}=
\begin{cases*}
\{\cdot,\cdot\} \quad \text{anticommutator if both variables are fermionic,} \\
[\cdot,\cdot] \quad \;\; \text{commutator otherwise.}
\end{cases*}
\end{equation*}}
\begin{equation}
[ C_\alpha, C_\beta \}=f_{\alpha\beta}{}^{\gamma}C_{\gamma}
\end{equation}
with some structure functions $f_{\alpha\beta}{}^{\gamma}$. The phase space action of such models depends on the particle spacetime coordinates $x^\mu$ together with ${\cal N}$ real fermionic superpartners $\Xi^\mu_I$ introduced to describe the spin degrees of freedom. In the case of a flat $d$-dimensional target spacetime, the action is given by\footnote{The Minkowski metric $\eta_{\mu\nu}\sim (-,+,\cdots,+)$ is used to raise and lower spacetime indices. Indices named $\mu,\nu,\dots$ refer to spacetime indices $(\mu,\nu=0,1,\dots,d-1)$, while those named $I,J,\dots$ stand for internal $SO(\N)$ indices $(I,J=1,\dots, \N)$.}
\begin{equation} \label{action}
S=\int d\tau \left[p_\mu\dot x^{\mu}+\frac{i}{2}\Xi_{\mu}\cdot\dot\Xi^{\mu}-e\,H-i\chi^I\,q_I-a^{IJ}\,J_{IJ}\right]\ , 
\end{equation}
where a dot indicates a contraction of the internal indices. The theory described by \eqref{action} should be regarded as a one-dimensional field theory living on the worldline, which provides a first-quantized description of relativistic particles of spin $s$. A few remarks regarding its field content are in order. \\ 

The canonical coordinates $(x^\mu, p_\mu, \Xi^\mu_I)\,$ upon quantization are subject to the canonical commutation relations (CCR)
\begin{equation} \label{CR}
[x^\mu, p_\nu]=i\,\delta^\mu_\nu\ ,\quad \{\Xi^\mu_I, \Xi^\nu_J\}=\delta_{IJ}\,\eta^{\mu\nu}\ . 
\end{equation}
The worldline supergravity multiplet in one dimension $(e,\chi^I, a_{IJ})$ contains the einbein $e$ which gauges worldline translations, the $\N$ gravitinos $\chi$ which gauge the worldline supersymmetry, and the gauge field $a$ for the symmetry which rotates by a phase the worldline fermions and gravitinos, the $R$-symmetry. In \eqref{action} they act as Lagrange multipliers for the suitable first-class constraints, which are the Hamiltonian $H$, the supercharges $q_I$, and the $R$-symmetry algebra generators $J_{IJ}$, choosing a specific order for the latter to avoid ambiguities
\begin{equation} \label{const}
H:=\frac{1}{2} \, p^\mu p_\mu\ , \quad q_I:=\Xi_I^\mu \, p_\mu\ , \quad J_{IJ}:=i\,\Xi_{[I}^\mu \, \Xi_{J]_\mu}\ .
\end{equation} 
Hamiltonian and supercharges together form the following one-dimensional algebra:
\begin{equation} \label{alg1}
\{q_I, q_J\}=2\,\delta_{IJ}\,H\ ,\quad [q_I, H]=0\ .
\end{equation}
These constraints have to be introduced to ensure the mass-shell condition and to eliminate negative norm states, ensuring the consistency of the model with unitarity at the quantum level \cite{Corradini:2015tik}. On the other hand, regarding the aforementioned $R$-symmetry algebra one finds:
\begin{align}
\begin{split} \label{alg2}
[J_{IJ}, q_K] &= i \left( \delta_{JK}q_I -\delta_{IK}q_J \right)\ , \\
[J_{IJ}, J_{KL}] &= i \left( \delta_{JK}J_{IL} -\delta_{IK}J_{JL} -\delta_{JL}J_{IK} +\delta_{IL}J_{JK} \right)\ .
\end{split}
\end{align}
The gauging of the $R$-symmetry group is optional and can be used to constrain the model to deliver pure spin $s$ states and have the minimal amount of degrees of freedom: to count them, one can construct the path integral on the one-dimensional torus of the free spinning particles, which has been achieved for all $\N$ in \cite{Bastianelli:2007pv}. Equations \eqref{alg1} together with \eqref{alg2} form the so-called $O(\N)$-\emph{extended worldline supersymmetry algebra}: it displays $\N$ supercharges $q_I$ which close on the Hamiltonian $H$ and which transform in the vector representation of $SO(\N)$, whose Lie algebra is described by the second line of \eqref{alg2}. In the following, we will refer only to \eqref{alg1} as ``$\N$ SUSY algebra", since it will be appropriate to distinguish the two algebras in light of the future BSRT procedure. It is rather convenient to work with a complex redefinition of the original fermionic variables $\Xi_I\to(\xi_i, \bar\xi^i)$, namely
\begin{equation}
\xi^\mu_i:=\tfrac{1}{\sqrt2}(\Xi^\mu_i+i\,\Xi^\mu_{i+2})\ , \quad \bar\xi^{\mu i}:=\tfrac{1}{\sqrt2}(\Xi^\mu_i-i\,\Xi^\mu_{i+2})\ , \quad \text{with} \; i=1,\dots,\tfrac{\N}{2}
\end{equation}
for the case of even $\N$, i.e. for particles with integer spin. The respective CCR \eqref{CR} become
\begin{equation}
\{\bar\xi^i_\mu, \xi_j^\nu\}=\delta^i_j\,\delta^\nu_\mu\ , \quad \{\bar\xi^i_\mu, \bar\xi^j_\nu\}=0=\{\xi_i^\mu, \xi_j^\nu\}\ . 
\end{equation}
Accordingly, both the supercharges and the $R$-symmetry generators split under the complex redefinition: we report the explicit expressions for the particle models of interest, namely the spin one and the spin two. By choosing a specific Fock vacuum, an arbitrary state $\ket{\Psi}$ in the Hilbert space is isomorphic to the wavefunction $\Psi(x,\xi)$, on which the conjugated momenta act as derivatives 
\begin{equation}
p_\mu=-i\partial_{\mu}\ , \quad \bar\xi_\mu=\frac{\partial}{\partial\xi^\mu}\ .
\end{equation}
Thus, for the case $\N=2$, the spin one model, one has the splitting $q_I \to (q, \bar q)$ and $J_{IJ} \to \mathrm{J}$ as follows
\begin{equation} \label{J}
\begin{array}{l}
q:=-i\,\xi^\mu\partial_\mu \\[2mm]
\bar{q}:=-i\,\partial^\mu\frac{\partial}{\partial\xi^\mu}
\end{array}
\ , \qquad \mathrm{J}:=\xi^\mu \frac{\partial}{\partial \xi^\mu} -\frac{d}{2}\ ,
\end{equation}
where the shift $-\tfrac{d}{2}$ in the definition of the $R$-symmetry generator is a quantum effect stemming from an antisymmetric ordering for the Grassmann variables \cite{Bastianelli:2005vk}. For the $\N=4$ case, the spin two, one has $q_I \to (q_i, \bar q^i)$ and the $SO(4)$-symmetry generators are realized maintaining manifest covariance only under a $u(2)$-subalgebra $J_{IJ} \to (\mathrm{J}_i{}^j, \mathrm{Tr}^{ij}, \mathrm{G}_{ij})$, explicitly realized as\footnote{In the following, whether a dot $\cdot$ indicates contractions on internal or spacetime indices should be clear within the context.}
\begin{equation} \label{JJ}
\begin{array}{l}
q_i:=-i\,\xi_i^\mu\partial_\mu \\[2mm]
\bar q^i:=-i\,\partial^\mu\frac{\partial}{\partial\xi_i^\mu} 
\end{array}
\ , \qquad 
\begin{array}{l}
\mathrm{J}_i{}^j:=\xi_i \cdot \frac{\partial}{\partial \xi_j} -\tfrac{d}{2}\, \delta_i^j \\[2mm]
\mathrm{Tr}^{ij}:= \frac{\partial^2}{\partial \xi_i \cdot \partial \xi_j} \\[2mm]
\mathrm{G}_{ij}:= \xi_i\cdot \xi_j
\end{array}\ .
\end{equation}
Note that the only surviving components of the $\mathrm{Tr}$ and $\mathrm{G}$ operators are those with $i \neq j$.

\section{BRST quantization} \label{sec2}
There are at least two methods for dealing with first-class constraints and constructing canonical quantization of gauge systems: the Dirac method and the BRST quantization. In the former approach, the procedure entails turning the constraints into operators $\hat{C}_\alpha$ and imposing that an element $\ket{\Psi}$ of the Hilbert space is physical if it gets annihilated by the constraint operators, i.e.
\begin{equation}
\hat{C}_\alpha\ket{\Psi}=0\ .
\end{equation}
This procedure has been extensively employed in \cite{Bastianelli:2008nm, Bastianelli:2014lia} to quantize both massless and massive spinning particle models. It is a relatively straightforward and powerful method, however, its application fails when the constraints algebra ceases to be first-class. These considerations eventually led to the BRST analysis of \cite{Dai:2008bh, Bonezzi:2018box, Bonezzi:2020jjq}, which is reviewed in the following to delineate the main steps of the quantization procedure, with the intention of applying them to the massive case. \\

Take the $\N=4$ spinning particle with a four-dimensional target space as an instructive example. The internal indices take values $i=1,2$ and the $\N=4$ worldline SUSY algebra is explicitly realized as
\begin{align} \label{alg}
\{q_i, \bar q^j\}=2\,\delta_{i}^{j}\,H\ ,\quad [q_i, H]=[\bar q^i, H]=\{q_i, q_j\}=\{\bar q^i, \bar q^j\}=0\ .
\end{align}
The initial step of the BRST procedure entails an enlargement of the Hilbert space to realize ghost-antighost pairs of operators $(g^\alpha, P_\alpha)$ associated with each constraint $C_\alpha$, with opposite Grassmann parity of the latter -- that is, anticommuting ghosts for bosonic constraints and commuting ghosts for fermionic constraints -- and canonical graded commutation relation
\begin{equation} \label{CIA}
[ P_\alpha,g^\beta \}=\delta^\beta_\alpha\ .
\end{equation}
Hence, we assign the fermionic pair $(c,b)$ to the Hamiltonian, and the bosonic superghost pairs $(\bar\gamma^i, \beta_i)$ and $(\gamma_i,\bar\beta^i)$ to the supercharges $q_i$ and $\bar q^i\,$ respectively, obeying
\begin{equation} \label{3.4}
\{b,c\}=1\ , \quad [\beta_i,\bar\gamma^j]=\delta_i^j\ , \quad [\bar\beta^j,\gamma_i]=\delta_i^j\ , 
\end{equation}
with ghost number assignments
\begin{align}
&{\gh(c,\gamma_i,\bar\gamma^i)=+1}\ , \\ 
&{\gh(b,\beta_i,\bar\beta^i)=-1}\ . 
\end{align}
The second step consists of constructing the BRST operator $\Q$. It is realized as follows
\begin{equation}\label{def}
\Q:=g^\alpha C_\alpha-\frac12(-1)^{\epsilon(g_\beta)} g^\beta g^\alpha f_{\alpha\beta}{}^{\gamma}P_\gamma\ ,
\end{equation}
where $\epsilon(g_\beta)$ is the Grassmann parity\footnote{We employ the following convention: the parity of $g^\alpha$ is $0$ if $g^\alpha$ is Grassmann even and $1$ if $g^\alpha$ is Grassmann odd.} of the ghost $g^\beta$. Equation \eqref{def} is exact if the structure functions $f_{\alpha\beta}{}^{\gamma}$ are constant, and can be derived by demanding the following properties for the BRST operator:
\begin{itemize}
\item It has to be anticommuting and of ghost number $+1$.
\item It has to act on the operators, corresponding to the original phase space variables prior to quantization, as the gauge transformations with the ghost variables replacing the gauge parameters. This, together with the latter requirement, is enough to constraint the structure to be $\Q=g^\alpha C_\alpha + \dots$
\item Finally, the BRST charge has to be nilpotent. This determines the second structure of \eqref{def} as can be checked by direct computation of $\{\Q,\Q\}$. 
\end{itemize}
Note that the BRST charge is nilpotent \emph{by construction} as long as the associated algebra is first-class. In more general cases, higher-order terms may appear and need to be determined by the nilpotency condition. Given its significance for future discussions, we shall show it for the case of the $\N=4$ spinning particle. The BRST operator associated with the first-class algebra \eqref{alg} is
\begin{equation} \label{QQ}
\Q= c\,H+\gamma_i\,\bar q^i+\bar\gamma^i\, q_i-2\bar\gamma^i\gamma_i\, b\ .
\end{equation}
To verify its nilpotency, it is first convenient to define $\bm{\nabla}:=\gamma \cdot \bar q+\bar\gamma \cdot q$, such that for any operator of the form \eqref{QQ} one always finds the following potential obstruction-terms to its nilpotency
\begin{equation} \label{sì}
\Q^2= -2 \, \bar\gamma \cdot\gamma\,H+\bm{\nabla}^2-2c\,\cancel{\left[H,\bm{\nabla}\right]}\ , 
\end{equation}
which are vanishing in the present case since 
\begin{equation} \label{no}
\bm{\nabla}^2=\gamma_i\bar \gamma^j \{\bar q^i, q_j \} \implies -2 \, \bar\gamma \cdot\gamma\,H+\bm{\nabla}^2=0\ .
\end{equation}
Notice the critical role played by the presence of the first-class algebra \eqref{alg} in both the cancellation of \eqref{sì} and of \eqref{no}. The same does not hold anymore in more general cases. In the next section, we will illustrate how to address this issue. \\

The careful reader might wonder about the absence of a set of ghosts associated with the $so(4)$ constraints. The key idea of the BRST procedure, as first developed in \cite{Dai:2008bh} providing a first-quantized description of the gluon, is to treat the $R$-symmetry constraints and the SUSY ones on different footings. The formers are imposed as constraints on the BRST Hilbert space, thus defining precisely the general dependence of the ``string field" $\Psi$ on the spacetime fields content.\footnote{Regarding terminology, we deliberately confuse ``string field" and ``BRST wavefunction" since $\Psi$ plays the same role in the worldline theory as it would in string field theory: its expansion as a linear combination of first-quantized states display coefficients which correspond to ordinary particle fields \cite{Gomis:1994he, Thorn:1988hm}.} It is within this restricted Hilbert space that the cohomology of the BRST charge has to be studied. The procedure remains consistent as long as the $R$-symmetry constraints \eqref{JJ} commute with the BRST charge: in order to achieve that, it is necessary to extend $(\mathrm{J}, \mathrm{Tr}, \mathrm{G})\to (\J, \gTr, \gG)$ as follows 
\begin{equation}
\begin{split}
\J_i{}^j &:= \xi_i\cdot\bar\xi^j+\gamma_i\bar\beta^j-\bar\gamma^j\beta_i-2\,\delta_i^j \ ,\\[2mm]
\gTr^{ij} &:= \bar\xi^i\cdot\bar\xi^j+\bar\gamma^i\bar\beta^j - \bar\gamma^j\bar\beta^i\ ,\\[2mm]
\gG_{ij} &:= \xi_i\cdot\xi_j+\gamma_i\beta_j - \gamma_j\beta_i\ .
\end{split} 
\end{equation}
The relevant $so(4)$ generators to be imposed as constraints on the BRST Hilbert space are the \emph{two number operators} $\J_i:=\J_i{}^i$ ($i$ not summed), namely the diagonal entries of $\J_i{}^j$, the \emph{Young antisymmetrizer} $\Y:=\J_1^2$, and finally the \emph{trace} $\gTr$, which implement the maximal reduction of the model. We collectively denote these constraints as $\mathcal{T}_\alpha:=(\J,\Y,\gTr)$. The BRST system is defined as follows
\begin{equation}\label{BRSTsystem}
\begin{split}
& \Q \, \Psi=0\;,\quad \delta\Psi=\Q \, \Lambda\;,\\[2mm]
&{\cal T}_\alpha \, \Psi=0\;,\quad {\cal T}_\alpha \, \Lambda=0\;,
\end{split} 
\end{equation}
and its consistency is ensured by $[\Q,\mathcal{T}_\alpha]=0$. This is indeed equivalent to saying that we are studying the cohomology of $\Q$ on the restricted Hilbert space of fixed $R$-charge defined by $\mathcal{H}_{\mathrm{red}}:=\ker\mathcal{T}_\alpha$. The system above will serve as the starting point for BRST quantization in both the massless and massive cases.


\subsection{Massless graviton on a flat spacetime}
The ghost vacuum $\ket{0}$ is chosen such that it is annihilated by $(b,\bar\gamma^i,\bar\beta^i)$ as in \cite{Bonezzi:2018box}, so that a general state $\ket{\Psi}$ in the BRST extended Hilbert space is isomorphic to the wavefunction $\Psi(x,\xi\,|\,c,\gamma,\beta)$, on which the antighosts act as derivatives, i.e. 
\begin{equation}
b=\frac{\partial}{\partial c}\ , \quad \bar\gamma^i=-\frac{\partial}{\partial \beta_i}\ , \quad \bar \beta^i=\frac{\partial}{\partial \gamma_i}\ .
\end{equation}
In the following, we collectively denote the ghost oscillators as $\g:=(c,\gamma,\beta)$. The BRST operator, making explicit the d'Alembertian $\Box=\eta^{\mu\nu}\partial_\mu\partial_\nu$ while adjusting the coefficients, acts as 
\begin{equation} \label{Q}
\Q=c\,\Box+\gamma_i\bar q^i-q_i\,\frac{\partial}{\partial\beta_i}-\gamma_i\,\frac{\partial^2}{\partial\beta_i\partial c}\ ,
\end{equation}
while, regarding the $so(4)$ generators
\begin{equation}
\begin{split} \label{rel}
\J_i &= N_{\xi_i}+N_{\gamma_i}+N_{\beta_i}-1\ , \\[2mm]
\Y &= \xi_1\cdot\frac{\partial}{\partial\xi_2}+\gamma_1\frac{\partial}{\partial\gamma_2}+\beta_1\frac{\partial}{\partial\beta_2}\ ,\\[2mm]
\gTr &= \frac{\partial^2}{\partial\xi_1\cdot\partial\xi_2}+\frac{\partial^2}{\partial\gamma_1\partial\beta_2}-\frac{\partial^2}{\partial\gamma_2\partial\beta_1}\ ,
\end{split} 
\end{equation}
where we have explicitly expressed the number operators, counting the number of oscillators with a fixed flavor index, and where we solved ambiguities in the definition of $\J_i$ by choosing a symmetric ordering for the ghosts. \\

It is useful for future reference to highlight the intermediate steps of the calculation. The first condition imposed by $\Psi \in \mathcal{H}_{\mathrm{red}}$, namely $\J_i \ket{\Psi}=0$, significantly reduces the components of $\Psi(x,\xi\,|\,\g)$ to
\begin{align} \label{massless}
\Psi(x,\xi\,|\,\g)&=
a_{\mu\nu}\,\xi^\mu_1\xi^\nu_2+b_{\mu}\,\xi_1^\mu\gamma_2+C_{\mu}\,\xi_1^\mu\beta_2+d_{\mu}\,\gamma_1\xi_2^\mu+e\,\gamma_1\beta_2+f_{\mu}\,\beta_1\xi_2^\mu \nonumber \\[1mm] 
&\phantom{=}+g\,\beta_1\gamma_2+k\,\gamma_1\gamma_2+l\,\beta_1\beta_2 \nonumber \\[1mm]
&\phantom{=}+a^{\ast}_{\mu\nu}\,\xi^\mu_1\xi^\nu_2c+b^{\ast}_{\mu}\,\xi_1^\mu\gamma_2c+C^{\ast}_{\mu}\,\xi_1^\mu\beta_2c+d^{\ast}_{\mu}\,\gamma_1\xi_2^\mu c+e^{\ast}\,\gamma_1\beta_2 c+f^{\ast}_{\mu}\,\beta_1\xi_2^\mu c \nonumber \\[1mm] 
&\phantom{=}+g^{\ast}\,\beta_1\gamma_2c+k^{\ast}\,\gamma_1\gamma_2c+l^{\ast}\,\beta_1\beta_2 c\ .
\end{align}
The latter expression is essentially a Taylor expansion in powers of $c$ taking the form $\Psi=\chi+\chi^{\ast}c$, where both terms $\chi$ and $\chi^{\ast}$ of the wavefunction contain oscillators and spacetime-dependent field components. Requiring the entire wavefunction to have fixed Grassmann parity and fixed ghost number forces the field components in $\chi^{\ast}$ to have opposite parities and ghost numbers decreased by one compared to those in $\chi$. It is important to note that we are not yet in the position to correctly identify (anti)fields: we first need to reduce $\Psi$ to include only the BV spectrum of the theory. Imposing the two remaining conditions we obtain
\begin{align}
\begin{split} \label{cond1}
&b_\mu =-d_\mu\ , \quad C_\mu=-f_\mu\ , \quad e=-g\ ,\quad k=0\ , \quad l=0\ , \\
&b^{\ast}_\mu =-d^{\ast}_\mu\ , \quad C^{\ast}_\mu=-f^{\ast}_\mu\ , \quad e^{\ast}=-g^{\ast}\ ,\quad k^{\ast}=0\ , \quad l^{\ast}=0\ ,
\end{split}
\end{align}
together with $a_{\mu\nu}=a_{\nu\mu}$ using $\Y \ket{\Psi}=0$, while from $\gTr \ket{\Psi}=0$ we get
\begin{equation} \label{cond2}
e=\tfrac{1}{2}a^\mu_\mu\ , \quad e^{\ast}=\tfrac{1}{2}a^{\ast \mu}_\mu\ .
\end{equation}
Now we can indeed interpret $\Psi$ as a spacetime BV string field, that contains the whole minimal BV spectrum of pure gravity along with auxiliary fields: we assign Grassman parity and ghost number to the component fields such that the entire wavefunction $\Psi$ has total even parity and ghost number zero. It becomes explicit with the following identifications for the fields
\begin{align} \label{renam}
a_{\mu\nu} \longrightarrow h_{\mu\nu} \quad a^{\mu}_{\mu} \longrightarrow h^{\mu}_{\mu}=:h \quad C^{\ast}_{\mu} \longrightarrow v_{\mu} \quad C_{\mu} \longrightarrow \zeta_{\mu}\ ,
\end{align}
and for the corresponding antifields
\begin{align}
a^{\ast}_{\mu\nu} \longrightarrow h^{\ast}_{\mu\nu} \quad b_{\mu} \longrightarrow v^{\ast}_{\mu} \quad b^{\ast}_{\mu} \longrightarrow \zeta^{\ast}_{\mu}\ .
\end{align}
Grassman parities and ghost numbers can be read from table \ref{table}. The most general state in $\mathcal{H}_{\mathrm{red}}$ is then
\begin{align} \label{BVstringfield}
\Psi(x,\xi\,|\,\g) &= h_{\mu\nu}(x)\,\xi^\mu_1\xi^\nu_2+\tfrac12\,h(x)\,(\gamma_1\beta_2-\gamma_2\beta_1)-\tfrac{i}{2}\,v_\mu(x)\,(\xi^\mu_1\beta_2-\xi^\mu_2\beta_1)c \nonumber\\[1mm]
&\phantom{=}-\tfrac{i}{2}\,\zeta_\mu(x)\,(\xi^\mu_1\beta_2-\xi^\mu_2\beta_1) \nonumber\\[2mm]
&\phantom{=}+h^{\ast}_{\mu\nu}(x)\,\xi^\mu_1\xi^\nu_2c+\tfrac12\,h^{\ast}(x)\,(\gamma_1\beta_2-\gamma_2\beta_1)c-\tfrac{i}{2}\,v^{\ast}_\mu(x)\,(\xi^\mu_1\gamma_2-\xi^\mu_2\gamma_1) \nonumber\\[1mm]
&\phantom{=}-\tfrac{i}{2}\,\zeta^{\ast}_\mu(x)\,(\xi^\mu_1\gamma_2-\xi^\mu_2\gamma_1)c\ . 
\end{align}
It contains the graviton $h_{\mu\nu}$ and its trace $h$, an auxiliary vector field $v_\mu$, and the diffeomorphism ghost $\zeta_\mu$, while the remaining components are the corresponding antifields.

\begin{table}[!ht]
\centering
\begin{tabular}{ |c|c|c|c| } 
 \hline
 BV role & Field & Grassmann parity & Ghost number \\
 \hline
 massless graviton & $h_{\mu\nu}$ & $0$ & $0$ \\
 trace & $h$ & $0$ & $0$ \\
 auxiliary vector & $v_\mu$ & $0$ & $0$ \\ 
 diffeomorphism ghost & $\zeta_\mu$ & $1$ & $1$ \\
 \hline
\end{tabular}
\caption{List of fields in the physical sector of the \emph{massless} $\N=4$ model with the corresponding ghost number and Grassmann parity.}
\label{table}
\end{table}
\vspace{1ex}
The final step involves evaluating the field equations of the theory, to verify that the model correctly reproduces a first-quantized representation of linearized gravity. This can be accomplished through the BRST closure equation
\begin{equation}
\Q\,\Psi(x,\xi\,|\,\g)=0
\end{equation} at ghost number zero, which produces the massless free spin two field equation
\begin{equation} \label{free}
\Box h_{\mu\nu}-2\,\partial_{(\mu}\partial\cdot h_{\nu)}+\partial_\mu\partial_\nu h =0\ , 
\end{equation}
and
\begin{equation} \label{free2}
\Box h -\partial^\mu \partial^\nu h_{\mu\nu}=0\ ,
\end{equation}
which implies a vanishing linearized Ricci scalar. To conclude this section, one is left to verify the accurate reproduction of the gauge symmetry: this is achieved from the ghost number zero part of
\begin{equation}
\delta\Psi=\Q\,\Lambda\ ,
\end{equation}
where $\Lambda \in \ker\mathcal{T}_\alpha$ contains the gauge parameters of the associated symmetry while having the same functional form as $\Psi$, with overall odd parity and ghost number $-1$, i.e. 
\begin{equation}\label{Lambdaparameter}
\Lambda=i\varepsilon_\mu(x)\,(\xi^\mu_1\beta_2-\xi^\mu_2\beta_1)+\cdots\ ,
\end{equation}
where the gauge parameter $\varepsilon_\mu$ has even parity and ghost number zero. As expected, the result is
\begin{equation}
\delta h_{\mu\nu}=2\partial_{(\mu} \varepsilon_{\nu)}\ .
\end{equation}
For future reference, the closure equation at ghost number one is reported below:
\begin{equation} \label{dimenticavo}
\Box \, \zeta_\mu=0\, \quad \partial^\mu \zeta_\mu=0\, \quad \partial_{(\mu} \, \zeta_{\nu)}=0\ .
\end{equation}
Equations \eqref{dimenticavo} represent a set of equations of motion for the diffeomorphism ghost $\zeta_\mu(x)$.

\subsection{Massless graviton on a curved background} \label{sec3.2}
The consistent coupling of the spinning particles to more general backgrounds beyond flat spacetime is a rather delicate matter \cite{Getzler:2016fek}. In previous works, a quantization \emph{à la} Dirac has been extensively employed. However, when the SUSY algebra fails to be first-class this method loses its validity, potentially leading to the misleading conclusion that quantization is not feasible in more general cases. This limitation was evident in the case of the $\N=4$ spinning particle, for which only certain restricted backgrounds were found to be viable until it was realized that the technique of BRST quantization offers a way to explore more general backgrounds, as recently discussed in \cite{Boffo:2022egz, Grigoriev:2021bes}. In this section, we review the exploration of such possibility. \\

The coupling of the massless $\N=4$ spinning particle to a curved background with metric $g_{\mu\nu}(x)$ is realized by the covariantization of the derivatives, i.e.
\begin{equation}
\hat \nabla_\mu:=\partial_\mu+\omega_{\mu\, ab}\,\xi^a\cdot\bar\xi^b\ ,\quad \text{with} \quad [\hat \nabla_\mu, \hat \nabla_\nu]=R_{\mu\nu\lambda\sigma}\,\xi^\lambda\cdot\bar\xi^\sigma\ , 
\end{equation}
where fermions carry flat Lorentz indices so that $\xi^\mu_i:=e^\mu_a(x)\,\xi^a_i$, introducing a background vielbein $e_\mu^a(x)$ and the torsion-free spin connection $\omega_{\mu\, ab}$. The covariant derivative operator $\hat \nabla_\mu$ reproduces the effect of the usual covariant (partial) derivative $\nabla_\mu$ ($\partial_\mu$) on the tensorial (scalar) components contained in the wavefunction \eqref{BVstringfield}, e.g.
\begin{equation}
\hat \nabla_\mu \Psi(x,\xi\,|\,\g) = \nabla_\mu h_{\alpha\beta}(x) \, \xi_1^\alpha\xi_2^\beta +\tfrac{1}{2} \partial_\mu h(x) \, (\gamma_1\beta_2-\gamma_2\beta_1)+\dots \ .
\end{equation}
The presence of a general background manifests itself as deformations of the original BRST system: indeed the supercharges become
\begin{equation}
\q_i:=-i\,\xi_i^a\,e^\mu_a\,\hat \nabla_\mu\ ,\quad \bar \q^i:=-i\,\bar\xi^{i\,a}\,e^\mu_a\,\hat \nabla_\mu\ .
\end{equation}
The main consequence is that the SUSY algebra does not close anymore:\footnote{As to not further burden the notation we denote $ \bm{R}_{\mu\nu}:=R_{\mu\nu\lambda\sigma}\,\xi^\lambda\cdot\bar\xi^\sigma$ and $\bm{R}:=R_{\mu\nu\lambda\sigma}\,\xi^\mu\cdot\bar\xi^\nu\,\xi^\lambda\cdot\bar\xi^\sigma$.}
\begin{align} \label{bo}
&\{\q_i, \q_j\}=-\xi^\mu_i\xi^\nu_j\, \bm{R}_{\mu\nu} \;,\quad \{\bar \q^i, \bar \q^j\}=-\bar\xi^{\mu\,i}\bar\xi^{\nu\,j}\, \bm{R}_{\mu\nu}\;,\quad \{\q_i, \bar \q^j\}=-\delta_i^j\nabla^2-\xi^\mu_i\bar\xi^{\nu\,j}\, \bm{R}_{\mu\nu}\ , \nonumber\\[3mm]
& [\nabla^2, \q_i]=i\,\xi^\mu_i\big(2\, \bm{R}_{\mu\nu}\,\hat \nabla^\nu-\nabla^\lambda \bm{R}_{\lambda\mu}-R_{\mu\nu}\hat \nabla^\nu\big)\ ,\nonumber\\[3mm]
& [\nabla^2, \bar \q^i]=i\,\bar\xi^{\mu\,i}\big(2\, \bm{R}_{\mu\nu}\,\hat \nabla^\nu-\nabla^\lambda \bm{R}_{\lambda\mu}-R_{\mu\nu}\hat \nabla^\nu\big)\ , 
\end{align}
where the Laplacian is defined as
\begin{equation} \label{laplace}
\nabla^2:=g^{\mu\nu}\hat \nabla_\mu \hat \nabla_\nu-g^{\mu\nu}\,\Gamma^\lambda_{\mu\nu}\,\hat \nabla_\lambda\ .
\end{equation}
The BRST operator needs to be deformed as well, requiring the construction of an ansatz due to the fact that the associated algebra is no longer first-class. The general approach involves considering the same general form \eqref{QQ}, but with the inclusion of more general terms accounting for the obstruction to the first-class character of the associated algebra, namely the curvature in the present case. Therefore, possible non-minimal couplings to the curvature, collectively denoted as $\Re$, must be incorporated inside the Hamiltonian. These couplings may act as obstructions to the nilpotency of the BRST operator, and the BRST analysis aims to determine which of these terms persist to ensure a nilpotent $\Q$. Avoiding higher powers of ghost momenta and assuming that derivatives are deformed only through minimal coupling, the ansatz for the BRST charge is
\begin{equation} \label{curvedQ}
\Q=c\,\D+\bm{\nabla}+\bar\gamma \cdot\gamma\,b\ ,
\end{equation}
where $\D$ is the deformed Hamiltonian in its operatorial form
\begin{equation} 
\D:=\nabla^2+\Re\ , \quad \text{with} \quad \Re:=R_{\mu\nu\lambda\sigma}\,\xi^\mu\cdot\bar\xi^\nu\,\xi^\lambda\cdot\bar\xi^\sigma+ \kappa R\ ,
\end{equation} 
with $\kappa$ a coefficient to be determined, and the $\bm{\nabla}$ operator has been conveniently redefined as 
\begin{equation}
\bm{\nabla}:=-i\,S^\mu\hat \nabla_\mu\ , \quad \text{with} \quad S^\mu:=\bar\gamma \cdot \xi^\mu+\gamma \cdot\bar\xi^{\mu}\ .
\end{equation}
Note that only the $so(4)$ constraints ${\cal T}_\alpha$ \eqref{rel} remain unchanged, therefore the wavefunction $\Psi(x,\xi\,|\,\g)$ is still expressed as in \eqref{BVstringfield}. \\

The BRST analysis starts with the general expression for $\Q^2$, namely the extension of its flat spacetime counterpart \eqref{sì} with adjusted coefficients:
\begin{equation} \label{Q22}
\Q^2=\bm{\nabla}^2+\bar\gamma \cdot\gamma\,\D+c\,\left[\D, \bm{\nabla}\right]\ , 
\end{equation}
which remains valid regardless of the specific spinning particle model coupled to a general curved background. In general, \eqref{Q22} includes two independent obstructions, which in this case are given by 
\begin{align} 
\bm{\nabla}^2+\bar\gamma \cdot\gamma\,\D &= -\tfrac12\,S^\mu S^\nu\, \bm{R}_{\mu\nu}+\bar\gamma \cdot\gamma\,\Re\ , \label{first obstruction} \\ 
\left[\D, \bm{\nabla}\right] &= -iS^\mu\nabla^\lambda \bm{R}_{\lambda \mu} +i S^\mu\nabla_\mu \bm{R}+\kappa [\,R,\bm{\nabla}]\ . \label{second obstruction}
\end{align}
Recall that the BRST cohomology is defined on the reduced Hilbert space $\mathcal{H}_{\mathrm{red}}$, that is, one needs to evaluate the obstructions on $\ker{\cal T}_\alpha$. Therefore, for the sake of the cohomology, it suffices to ensure nilpotency of the BRST charge \emph{when acting} on the physical sector of the theory:
\begin{equation}
 \Q^2\stackrel{{\ker}{\cal T}_\alpha}{=}0 \quad \text{i.e.} \quad \Q^2\Psi(x,\xi\,|\,\g)=0\ .
\end{equation}
The effect is for instance that any contribution with at least three barred oscillators is set to zero. Equations \eqref{first obstruction}--\eqref{second obstruction} get then reduced to
\begin{align}
\bm{\nabla}^2+\bar\gamma \cdot\gamma\,\D &\stackrel{{\ker}{\cal T}_\alpha}{=}-\,\bar\gamma\cdot\gamma\left(2\,R_{\mu\nu}\,\xi^\mu \cdot\bar\xi^\nu-\kappa \, R\right)\ , \label{first-first} \\
\begin{split} \label{second-second}
\left[\D, \bm{\nabla}\right] &\stackrel{\ker{\cal T}_\alpha}{=}-iS^\mu\nabla^\lambda \bm{R}_{\lambda\mu}+i \big(2 \nabla^\lambda \bm{R}_{\lambda
\mu}\gamma\cdot\bar\xi^\mu-S^\mu\nabla_\mu R_{\nu\lambda}\xi^\nu \cdot\bar\xi^\lambda\big)\\
&\phantom{\stackrel{\ker{\cal T}_\alpha}{=}}\;+i\kappa \,S^\mu\nabla_\mu R\ .
\end{split}
\end{align}
The conclusion is that the nilpotency of the BRST charge, in the massless case, is achieved only on Einstein manifolds, i.e.
\begin{equation}
\Q^2 \Psi(x,\xi\,|\,\g)=0 \; \iff \; R_{\mu\nu}=\lambda\,g_{\mu\nu}\ ,
\end{equation}
upon setting $\kappa=\tfrac{1}{2}$. Let us clarify this crucial point for the upcoming discussion: while the obstruction given in \eqref{second-second} vanishes by itself on Einstein spaces, the one in \eqref{first-first} is zero \emph{only} when acting on the physical sector:
\begin{equation}
\Q^2 \Psi(x,\xi\,|\,\g) \ni -2\lambda\,\bar\gamma\cdot\gamma\left(\xi^\mu \cdot\bar\xi_\mu-1\right) \, \Psi(x,\xi\,|\,\g)=0\ .
\end{equation}
The closure equation evaluated using the deformed BRST charge yields the correct field equations for a massless\footnote{It is a rather fascinating topic the concept of mass on a general spacetime, which is not well defined. In particular, one should be careful to call a particle \emph{massless}: see \cite{Deser:1983mm} for a discussion on the connection between gauge invariance, masslessness, and null cone propagation.} graviton on Einstein spaces, namely
\begin{align}
\nabla^2h_{\mu\nu}-2\nabla_{(\mu}\nabla \cdot h_{\nu)}+\nabla_\mu\nabla_\nu h+2 R_{\mu\alpha\nu\beta}h^{\alpha\beta} =0\ ,
\end{align}
together with
\begin{align}
( \nabla^2 +\lambda )h-\nabla^\mu\nabla^\nu h_{\mu\nu}=0\ .
\end{align}
The correct gauge symmetry $\delta h_{\mu\nu}=2\nabla_{(\mu}\varepsilon_{\nu)}$ is obtained as well.


\section{Giving mass to the graviton} \label{sec3}
In this section, we present two methods to confer a mass to a spinning particle model. The first one, the Scherk-Schwarz mechanism, is analogous to the Kaluza-Klein compactification. In the second method, the auxiliary oscillators approach, the model is treated as a truncation of the RNS open superstring \cite{Green:2012pqa}. We employ both methods to give mass to the graviton on a flat background and subsequently discuss the results. However, before delving into the graviton case, we begin by examining the $\N=2$ scenario, in order to introduce the techniques and investigate the role of the mass as a potential obstruction to nilpotency of the BRST charge. 

\subsection{$\N=2$ massive spinning particle and the Proca Theory}
The massless $\N=2$ spinning particle has been thoroughly examined in previous works, leading to a first-quantized description of massless antisymmetric tensor fields of arbitrary rank \cite{Berezin:1976eg, Gershun:1979fb, Howe:1988ft, Howe:1989vn, Brink:1976uf}. It has been shown how to provide a first-quantized description of the photon both on a flat and on a curved spacetime \cite{Bastianelli:2005vk, Grigoriev:2021bes}. Additionally, the coupling to a non-abelian background has been explored, reproducing the non-linear Yang-Mills equations \cite{Dai:2008bh} upon BRST quantization. \\

The massive scenario has undergone investigation as well: the procedure \emph{à la} Kaluza-Klein has been implemented in \cite{Brink:1976uf, Bastianelli:2005uy} where it has been shown how to embed the model in an \emph{arbitrary curved} spacetime background. This has allowed for the utilization of this model in the context of worldline quantum field theory formalism, particularly in modeling classical scatterings of compact objects in general relativity (see \cite{Mogull:2020sak, Jakobsen:2021zvh} and related literature). More recently, the coupling to both an abelian and a non-abelian vector background field has been investigated in \cite{Carosi:2021wbi}, where the authors first introduced the auxiliary oscillators approach.

\subsubsection{Mass improvement on flat spacetime}

The main features of the $\N=2$ model can be immediately derived from section \ref{sec1} by setting $I=1,2$. For a comprehensive analysis of the action, the first-class algebra, and the canonical quantization, we direct the reader to \cite{Bastianelli:2005uy}. In this work, we will provide a concise overview of the essential elements of the BRST quantization of the massive case on flat and, most notably, on curved background as a toy model for the subsequent analysis, which is not covered in the literature mentioned above. \\

The procedure consists of starting with the massless theory formulated in one dimension higher, and subsequently reducing the dimensionality of the target space through the introduction of suitable constraints \cite{Bastianelli:2014lia}. We consider the model to live in a flat spacetime of the form $\mathcal{M}_d\times S^1$ of $D=d+1$ dimensions with coordinates and worldline superpartners given by
\begin{align}
x^M = (x^\mu , x^D)\ , \quad \xi^M=(\psi^\mu,\theta)\ , \quad \bar \xi^{M}=(\bar \psi^{\mu},\bar \theta)\ ,
\end{align}
with the index splitting $M=(\mu, D)$. The idea is to gauge the compact direction $x^D$, corresponding to $S^1$, by imposing the first-class constraint 
\begin{equation}
p_D-m=0\ .
\end{equation}
This results in the phase space action \eqref{action} having a leftover term $\sim \dot x^D$, which can be regarded as a total derivative and thus dropped. The CCR \eqref{CR} are realized explicitly as
\begin{equation} 
[x^\mu, p_\nu]=i\,\delta^\mu_\nu\ ,\quad \{\bar \psi^\mu, \psi^\nu\}=\eta^{\mu\nu}\ , \quad \{\bar \theta, \theta\}=1\ , 
\end{equation}
with the other (anti)commutators being zero. The $\N=2$ SUSY constraints get modified by the presence of the mass, taking the form
\begin{align}
H=\frac12 \left(p^\mu p_\mu+ m^2\right)\ , \quad
q=p_\mu \psi^\mu + m \theta\ , \quad 
\bar q=p_\mu \bar \psi^\mu + m \bar \theta\ ,
\end{align}
while the $R$-symmetry generator \eqref{J} becomes
\begin{equation}
\mathrm{J} =\psi^\mu \bar \psi_\mu +\theta \bar \theta-\frac{d+1}{2}\ .
\end{equation}
Together they still satisfy the same $\N=2$ supersymmetry algebra despite the mass improvement
\begin{align} \label{boo}
\{q, \bar q\}=2H\ ,\quad [q, H]=[\bar q, H]=\{q, q\}=\{\bar q, \bar q\}=0\ ,
\end{align}
which successfully remains first-class. Note the key role played by the surviving fermionic coordinates coming from the extra dimension $\theta:=\xi^D$ and $\bar \theta:=\bar \xi^{D}$, which are responsible for the introduction of the mass term into the theory. Indeed, in the limit where $(\theta, \bar \theta) \to 0$ the massless theory must be recovered as a consistency check. \\

At this point, the BRST quantization on a four-dimensional flat spacetime proceeds smoothly, just as described for the $\N=4$ case. Upon enlarging the Hilbert space to realize the set of ghost operators $(c, \bar \gamma, \gamma)$ with relative momenta $(b,\beta, \bar \beta)$ as in \eqref{3.4}, the BRST charge is constructed as usual
\begin{equation}
\Q= c\,H+\gamma\,\bar q+\bar\gamma\, q-2\bar\gamma\gamma\, b\ ,
\end{equation}
and the $so(2)$ constraint has to be extended to include ghost contributions as follows
\begin{equation}
\J:= \psi^\mu \bar \psi_\mu +\theta \bar \theta +\gamma\bar\beta-\bar\gamma\beta-\tfrac{3}{2}\ ,
\end{equation}
with the usual prescription employed to resolve ambiguities. Nilpotency of $\Q$ is still guaranteed, despite the mass improvement, since the associated algebra is first-class
\begin{equation}
\Q^2=0\ ,
\end{equation}
and the BRST system is consistent since $[\Q,\J]=0$. The physical sector is defined as the eigenspace of $\J$ with a fixed $R$-charge $-\tfrac{1}{2}$,\footnote{This condition is equivalent to demanding that $\Psi \in{\ker}{\tilde \J}$, with $\tilde \J$ being the shifted $SO(2)$ constraint $\tilde \J = \J+\tfrac{1}{2}$. The procedure remains consistent as long as $[\Q,\tilde \J]=0$. Eventually, the shift can be interpreted as the introduction of a Chern-Simons term selecting the desired degrees of freedom, as in \cite{Bastianelli:2005uy}.} i.e. physical states are eigenstates $\J\ket{\Psi}=-\tfrac{1}{2}\ket{\Psi}$ and are isomorphic to wavefunctions
\begin{align}
\begin{split}
\Psi(x,\psi,\theta\,|\,\g) &= A_\mu(x)\psi^\mu-i\varphi(x)\theta +\phi(x) \beta c +\Phi(x) \beta\\[1mm]
&\phantom{=}+A_\mu
^{\ast}(x) \psi^\mu c-i\varphi^{\ast}(x) \theta c+\phi^{\ast}(x)\gamma+\Phi^{\ast}(x)\gamma c\ .
\end{split}
\end{align}
Requiring $\Psi$ to be Grassmann-odd and have ghost number zero, it can be interpreted as a spacetime BV string field displaying the complete minimal BV spectrum of the Proca theory along with an auxiliary field. The Grassmann parities and ghost numbers of the components can be found in table \ref{table3}. In contrast to the massless case, the physical fields include not only the massive spin one $A_\mu$ but also the St\"uckelberg scalar $\varphi$. The inclusion of the St\"uckelberg scalar becomes necessary to restore the $U(1)$ gauge symmetry \cite{Stueckelberg:1957zz}, which is broken due to the introduction of the mass. Notably, the spectrum also includes the associated scalar ghost $\Phi$ in the spectrum.

\begin{table}[!ht]
\centering
\begin{tabular}{ |c|c|c|c| } 
 \hline
 BV role & Field & Grassmann parity & Ghost number \\
 \hline
 massive spin one & $A_{\mu}$ & $0$ & $0$ \\
 St\"uckelberg scalar & $\varphi$ & $0$ & $0$ \\
 auxiliary scalar & $\phi$ & $0$ & $0$ \\ 
 scalar ghost & $\Phi$ & $1$ & $1$ \\
 \hline
\end{tabular}
\caption{List of fields in the physical sector of the \emph{massive} $\N=2$ model with the corresponding ghost number and Grassmann parity.}
\label{table3}
\end{table}
\vspace{1ex}
The field equations, upon solving for the auxiliary field, are
\begin{align}
\left( \Box-m^2 \right)A_\mu-\partial_\mu\partial \cdot A-m\partial_\mu\varphi&=0\ , \\
\Box \varphi+m\partial_\mu A^\mu&=0\ , 
\end{align}
which are indeed the Proca field equations with the scalar $\varphi(x)$ playing the role of St\"uckelberg field, while the gauge symmetry, from $\delta \Psi=\Q \Lambda$ with
\begin{equation} \label{gg}
\Lambda=i\Sigma(x) \beta +\cdots\ ,
\end{equation}
is
\begin{equation}
\delta A_\mu=\partial_\mu\Sigma\ , \quad \delta\varphi=-m \, \Sigma\ ,
\end{equation}
where $\Sigma(x)$ is a local gauge parameter, of even parity and with ghost number zero. The same conclusions hold when employing the auxiliary oscillators approach, although it is not explicitly shown here. The subsequent analysis will elaborate on the limitations of this approach in the context of the graviton case. 

\subsubsection{Mass obstruction on curved spacetime}
We are finally in the position to investigate whether the presence of the mass plays a role when a curved background is considered, generalizing the massless BRST analysis of \cite{Grigoriev:2021bes}. The coupling of the massive $\N=2$ spinning particle to gravity is realized by the covariantization of the reduced model just discussed, with the mass already present in the theory. The supercharges are then deformed as follows
\begin{equation}
\q:=-i\,\psi^a\,e^\mu_a\,\hat \nabla_\mu\ +m\theta\ ,\quad \bar \q:=-i\,\bar\psi^{a}\,e^\mu_a\,\hat \nabla_\mu+m\bar\theta\ .
\end{equation}
Unlike the situation with the $\N=4$ spinning particle \eqref{bo}, the $\N=2$ algebra \eqref{boo} remains first-class regardless of a particular background, upon a suitable redefinition of the Hamiltonian 
\begin{equation} \label{H}
\D:=\nabla^2-m^2+\bm{R}\ .
\end{equation}
Indeed, the SUSY algebra takes the form
\begin{align} 
\begin{split} \label{ALG}
\{\q, \bar \q\}=-\D\ , \quad \{\q, \q\}=\{\bar \q, \bar \q\}&=0\ , \\
[\q,\D]=[\bar \q,\D]&=0\ ,
\end{split}
\end{align}
where the vanishing of the top-right commutators is guaranteed by the cyclic identity for the Riemann tensor. The corresponding BRST operator is defined as usual $\Q=c\,\D+\bm{\nabla}+\bar\gamma \gamma\,b$, with potential obstructions incorporated into the definition of $\bm{\nabla}$ with respect to the massless case:
\begin{equation}
\bm{\nabla}:=-i\,S^\mu\hat \nabla_\mu+ m\rho\ , \quad \rho:=\bar \gamma \, \theta+\gamma \, \bar \theta\ .
\end{equation}

The existence of an associated first-class algebra \eqref{ALG} ensures the nilpotency of the BRST operator without any further conditions on the background metric,\footnote{Actually, the only requirement is that of a torsionless connection, i.e. $T^\mu_{\nu\rho}:=\Gamma^\mu_{[\nu\rho]}=0$.} which is expected from QFT considerations. While this has been already established for the massless case, the same holds even in the massive one, as the mass does not affect the algebra, as we now shall work out explicitly. Starting from the general expression
\begin{equation}
\Q^2=\bm{\nabla}^2+\bar\gamma \gamma\,\D+c\,\left[\D, \bm{\nabla}\right]\ ,
\end{equation}
one finds the following independent obstructions 
\begin{align} 
\bm{\nabla}^2+\bar\gamma \gamma\,\D &=-\frac{1}{2}S^\mu S^\nu \bm{R}_{\mu\nu}+\cancel{\bar\gamma \gamma \, m^2}+ \bar\gamma \gamma \left(\cancel{-m^2}+ \bm{R}\right)\ , \label{uno} \\
\left[\D, \bm{\nabla}\right] &=-iS^\mu
\nabla^\lambda\bm{R}_{\lambda\mu}+iS^\mu
\nabla_\mu\bm{R}\ .\label{due}
\end{align}
In the present case, evaluating the BRST cohomology on the reduced Hilbert space, $\ker\mathcal{\tilde \J}$, has the effect of setting to zero any obstruction of the form $\mathcal{O}^{AB}\bar Z_A\bar Z_B$ with $\mathcal{O}^{AB}$ arbitrary operators since an arbitrary state in the physical sector has the form
\begin{equation}
\Upsilon_{A}(x)\,Z^A + \Omega_{B}(x)\,Z^B\,c \quad \text{for} \quad Z^A=(\psi^\mu, \theta, \gamma,\beta)\ .
\end{equation}
This is enough to conclude that indeed
\begin{equation}
\Q^2\stackrel{\ker\mathcal{\tilde \J}}{=}0 
\end{equation}
without further conditions on the background; in other words, the massive $\N=2$ spinning particle can be coupled to off-shell gravity. Remarkably, the mass plays no role in the BRST algebra, as it is trivially canceled in \eqref{uno} and commutes with any operator inside \eqref{due} being constant. It is worth noting that introducing a term proportional to the Ricci curvature inside the Hamiltonian \eqref{H} $\D \to \D +\kappa R$, would force $\kappa$ to be zero (either that or a vanishing Ricci scalar $R=0$). This is dictated by the BRST algebra \eqref{uno}--\eqref{due}, as any term of that form is incompatible with the closure of the $\N=2$ SUSY algebra \eqref{boo}. Interestingly, the $\N=2$ spinning particle appears to select only the minimal coupling to the background, as we shall check from the equations of motion. 
From the closure equation $\Q\Psi=0$, one finds
\begin{align}
( \nabla^2-m^2 )A_\mu-R_{\mu\nu}A^\nu
+\nabla_\mu \phi&=0\ , \\
( \nabla^2-m^2 )\varphi-i m\phi&=0\ , \\
\phi+i m\varphi-i \nabla_\mu A^\mu&=0\ ,
\end{align}
which, upon eliminating the auxiliary field, yield
\begin{align}
( \nabla^2-m^2)A_\mu-\nabla_\mu\nabla_\nu A^\nu+R_{\mu\nu}A^\nu-m\nabla_\mu\varphi&=0\ , \label{proca1} \\
\nabla^2\varphi+m\nabla_\mu A^\mu&=0\ . \label{proca2}
\end{align}
Equations \eqref{proca1}--\eqref{proca2} represent the field equations of the Proca theory on a general curved spacetime in its St\"uckelberg formulation. In particular, the $\N=2$ spinning particle provides a first-quantized version of the minimal extension for a theory of a free massive vector field in curved spacetime, without the inclusion of possible non-minimal couplings to the background in the corresponding spacetime quantum field theory \cite{Belokogne:2015etf, Buchbinder:2017zaa}. Their absence is attributed to the reasons previously discussed. As a final note, the gauge symmetry, using \eqref{gg}, reads
\begin{align}
\delta A_\mu=\nabla_\mu\Sigma\ , \quad \delta\varphi=-m \, \Sigma\ .
\end{align}

\subsection{$\N=4$ massive spinning particle and Linearized Massive Gravity}
In the remainder of this section, the methods previously discussed are exploited to give a mass to the graviton, considering a flat target spacetime for the time being.

\subsubsection{Dimensional reduction approach}
The first method proceeds along the lines of what has been shown for the $\N=2$ case, namely through the reduction of the higher-dimensional massless model. The fermionic coordinates carry a flavor index 
\begin{align}
\xi^M_i=(\psi^\mu_i,\theta_i)\ , \quad \bar \xi^{M i}=(\bar \psi^{\mu i},\bar \theta^i)\ ,
\end{align}
with CCR
\begin{equation} 
\{\bar \psi^{\mu i}, \psi^{\nu}_j\}=\delta^i_j\,\eta^{\mu\nu}\ , \quad \{\bar \theta^i, \theta_j\}=\delta^i_j\ .
\end{equation}
The same applies to the SUSY constraints \eqref{const}
\begin{align}
H=\frac12 \left(p^\mu p_\mu+ m^2\right)\ ,\quad q_i=p \cdot \psi_i + m \theta_i\ , \quad \bar q^i=p \cdot \bar \psi^i + m \bar \theta^i\ ,
\end{align}
that satisfy the same first-class algebra 
\begin{align}
\{q_i, \bar q^j\}=2\delta_i^j H\ ,\quad [q_i, H]=[\bar q^i, H]=\{q_i, q_j\}=\{\bar q^i, \bar q^j\}=0
\end{align}
despite the mass improvement. Regarding the $so(4)$ symmetry constraints, equations \eqref{JJ} become
\begin{align}
\begin{split}
\mathrm{J}_i{}^j &=\psi_i \cdot \frac{\partial}{\partial \psi_j} +\theta_i \; \frac{\partial}{\partial \theta_j} -\frac{d+1}{2}\, \delta_i^j\ , \\
\mathrm{Tr}^{ij} &= \frac{\partial^2}{\partial \psi_i \cdot \partial \psi_j}+\frac{\partial^2}{\partial \theta_i \; \partial \theta_j}\ , \\
\mathrm{G}_{ij} &= \psi_i\cdot \psi_j+\theta_i \; \theta_j\ .
\end{split}
\end{align}
At this point, the BRST quantization proceeds as outlined in section \ref{sec2}, with the BRST system defined by \eqref{BRSTsystem}. The BRST operator \eqref{Q} takes the usual form when acting on wavefunctions $\Psi(x,\psi,\theta\,|\,\g)$
\begin{equation} \label{Q2}
\Q=c\,(\Box-m^2)+\gamma_i\,\bar q^i-q_i\,\frac{\partial}{\partial\beta_i}-\gamma_i\,\frac{\partial^2}{\partial\beta_i\partial c}
\end{equation}
and is still nilpotent, for the same reasons outlined in the previous section. The relevant $so(4)$ generators \eqref{rel} to be imposed on the extended BRST Hilbert space are extended to commute with $\Q$, including the extra fermionic oscillators
\begin{equation} \label{Ta}
\begin{split}
\J_i &= N_{\psi_i}+N_{\theta_i}+N_{\gamma_i}+N_{\beta_i}-\frac{d-1}{2}\ , \\[2mm]
\Y &= \psi_1\cdot\frac{\partial}{\partial\psi_2}+\theta_1 \; \frac{\partial}{\partial\theta_2}+\gamma_1\frac{\partial}{\partial\gamma_2}+\beta_1\frac{\partial}{\partial\beta_2}\ ,\\[2mm]
\gTr &= \frac{\partial^2}{\partial\psi_1\cdot\partial\psi_2}+\frac{\partial^2}{\partial\theta_1 \; \partial\theta_2}+\frac{\partial^2}{\partial\gamma_1\partial\beta_2}-\frac{\partial^2}{\partial\gamma_2\partial\beta_1}\ ,
\end{split} 
\end{equation}
The physical subspace has a fixed $U(1)\times U(1)$ charge corresponding to $\tfrac{3-d}{2}$. Let us comment on the fact that only in three spacetime dimensions the wavefunction $\Psi \in \ker\J_i$, which becomes relevant when dealing with the construction of the worldline path integral.\footnote{This analysis has been conducted in order to correctly calculate the one-loop divergencies of the effective action of linearized massive gravity and will be published in a separate paper \cite{Fecit:2024}.} In four spacetime dimensions, physical states are charged $-\tfrac{1}{2}$, and we shall consider the shifted operator $\tilde \J = \J+\tfrac{1}{2}$ for the sake of notational simplicity. To emphasize the impact of the introduction of a mass term, we follow the same steps as in the massless case to impose the condition $\Psi \in \mathcal{H}_{\mathrm{red}}$, with the reduced Hilbert space defined as the kernel of $\tilde \J_i$ and $\Y, \gTr$ \eqref{Ta}, still collectively denoted as $\mathcal{T}_\alpha$\color{black}. From $\tilde \J_i \ket{\Psi}=0$ descends the general form of the wavefunction
\begin{align} \label{comp}
\Psi(x,\psi,\theta\,|\,\g)&=\left. \Psi\right|_{m=0} \nonumber \\[1mm]
&\phantom{=}+n\,\theta_1\gamma_2+p\,\theta_1\beta_2+q\,\gamma_1\theta_2+r\,\beta_1\theta_2+s_\mu\,\theta_1\psi_2^\mu+t_\mu\,\psi_1^\mu\theta_2 +u\,\theta_1\theta_2 \\[1mm]
&\phantom{=}+n^{\ast}\,\theta_1\gamma_2c+p^{\ast}\,\theta_1\beta_2c+q^{\ast}\,\gamma_1\theta_2c+r^{\ast}\,\beta_1\theta_2c+s^{\ast}_\mu\,\theta_1\psi_2^\mu c+t^{\ast}_\mu\,\psi_1^\mu\theta_2c+u^{\ast}\,\theta_1\theta_2c\ , \nonumber
\end{align}
where in the first line $\left. \Psi\right|_{m=0}$ denotes the massless wavefunction \eqref{massless}. The condition $\Y \ket{\Psi}=0$ produces, in addition to the ``massless" contribution \eqref{cond1},
\begin{align}
\begin{split}
&n=-q\ , \quad p=-r\ , \quad s_\mu=t_\mu\ , \\
&n^{\ast}=-q^{\ast}\ , \quad p^{\ast}=-r^{\ast}\ , \quad s^{\ast}_\mu=t^{\ast}_\mu\ ,
\end{split}
\end{align}
while the remaining $\gTr \ket{\Psi}=0$ produces
\begin{align}
e=\tfrac{1}{2}a^\mu_\mu+\tfrac{1}{2}u\ , \quad e^{\ast}=\tfrac{1}{2}a^{\ast \mu}_\mu+\tfrac{1}{2}u^{\ast}\ .
\end{align}
It is now possible to identify the BV spectrum of the theory, which becomes evident after renaming the field components as in \eqref{renam} along with 
\begin{align} \label{renam2}
s_{\mu} \longrightarrow A_{\mu} \quad u \longrightarrow \varphi \quad p \longrightarrow \phi \quad p^{\ast} \longrightarrow \Phi
\end{align}
and, for the corresponding antifields,
\begin{align}
s^{\ast}_{\mu} \longrightarrow A^{\ast}_{\mu} \quad u^{\ast} \longrightarrow \varphi^{\ast} \quad n \longrightarrow \phi^{\ast} \quad n^{\ast} \longrightarrow \Phi^{\ast}\ .
\end{align}
Requiring $\Psi$ to be Grassmann-even and have ghost number zero the field content gets assigned with the corresponding Grassman parities and ghost numbers as reported in table \ref{table2}. The most general string field $\Psi$ in $\ker\mathcal{T}_\alpha$ reads
\begin{align} \label{BVstringfield2}
\Psi(x,\psi,\theta\,|\,\g) &= h_{\mu\nu}(x)\,\psi^\mu_1\psi^\nu_2+\tfrac12\,h(x)\,(\gamma_1\beta_2-\gamma_2\beta_1)-\tfrac{i}{2}\,v_\mu(x)\,(\psi^\mu_1\beta_2-\psi^\mu_2\beta_1)c \nonumber\\[1mm]
&\phantom{=}-\tfrac{i}{2}\,\zeta_\mu(x)\,(\psi^\mu_1\beta_2-\psi^\mu_2\beta_1)\nonumber\\[2mm]
&\phantom{=}-iA_\mu(x)\,(\theta_1\psi_2^\mu+\psi_1^\mu\theta_2)-\varphi(x)\,(2\theta_1\theta_2+\gamma_1\beta_2-\gamma_2\beta_1)+\phi(x)\,(\theta_1\beta_2-\theta_2\beta_1)c\nonumber\\[1mm]
&\phantom{=}+\Phi(x)\,(\theta_1\beta_2-\theta_2\beta_1)\ ,
\end{align}
where, for the sake of simplicity, the antifields content is left implicit.

\begin{table}[!ht]
\centering
\begin{tabular}{ |c|c|c|c| } 
 \hline
 BV role & Field & Grassmann parity & Ghost number \\
 \hline
 massless graviton & $h_{\mu\nu}$ & $0$ & $0$ \\
 trace & $h$ & $0$ & $0$ \\
 auxiliary vector & $v_\mu$ & $0$ & $0$ \\ 
 diffeomorphism ghost & $\zeta_\mu$ & $1$ & $1$ \\
 St\"uckelberg vector & $A_\mu$ & $0$ & $0$ \\
 St\"uckelberg scalar & $\varphi$ & $0$ & $0$ \\
 auxiliary scalar & $\phi$ & $0$ & $0$ \\
 scalar ghost & $\Phi$ & $1$ & $1$ \\
 \hline
\end{tabular}
\caption{List of fields in the physical sector of the \emph{massive} $\N=4$ model with the corresponding ghost number and Grassmann parity in the dimensional reduction approach.}
\label{table2}
\end{table}
\vspace{1ex}
A few comments are in order. Firstly, it can be verified that the first two lines of \eqref{BVstringfield2} correspond to the field content in the massless case \eqref{BVstringfield}. This is expected since, as previously suggested,
\begin{equation}
\Psi \xrightarrow[]{(\theta, \bar \theta) \to 0}\left. \Psi \right|_{m=0}
\end{equation}
up to an obvious redefinition of the scalar field $\varphi(x)$ which can be absorbed into the trace $h(x)$. Taking into account also the other two lines, it is clear that $\Psi$ contains the whole minimal BV spectrum of LMG, with the inclusion, with respect to the massless spectrum, of the two St\"uckelberg fields $A_\mu$ and $\varphi$, a scalar ghost field $\Phi$ and an auxiliary scalar field $\phi$ \cite{Boulanger:2018dau}.

At this point, we proceed to analyze the field equations, which are derived from the closure equation at ghost number zero using the BRST operator \eqref{Q2}. The result is
\begin{subequations} \label{eq}
\begin{align}
\left(\Box-m^2\right)h_{\mu\nu}-\partial_{(\mu}v_{\nu)}&=0 \label{eq1}\ , \\
\left(\Box-m^2\right)h-2\left(\Box-m^2\right)\varphi-\partial \cdot v+2m\phi&=0\ , \label{eq5}\\
\left(\Box-m^2\right)A_\mu +\partial_\mu \phi +\tfrac{m}{2} v_\mu
&=0\ , \\
\left(\Box-m^2\right)\varphi-m\phi&=0\ , \label{eq3}\\
v_\mu -2\partial \cdot h_\mu +\partial_\mu h-2m A_\mu -2\partial_\mu \varphi&=0 \label{eq4}\ , \\
\partial \cdot A +m\varphi +\tfrac{1}{2}mh+\phi&=0\label{eq2}\ .
\end{align}
\end{subequations}
It is possible to solve \eqref{eq4} and \eqref{eq2} for the auxiliary fields $v_\mu$ and $\phi$ respectively, leading to the following equations
\begin{subequations} \label{main}
\begin{align}
&\left(\Box-m^2\right)h_{\mu\nu}-2\partial_{(\mu} \partial \cdot h_{\nu)}+\partial_\mu\partial_\nu h=2m\partial_{(\mu}A_{\nu)}+2\partial_\mu \partial_\nu \varphi\ ,\label{main2.1} \\
&\left(\Box-m^2\right)h-\partial^\mu\partial^\nu h_{\mu\nu}=2m \partial \cdot A+2\Box\varphi\ , \label{main2.2}\\
&\phantom{(}\Box A_\mu -\partial_\mu \partial \cdot A=m\left( \partial_\mu h-\partial \cdot h_\mu \right)\ ,\label{main2.3} \\
&\phantom{(}\Box\varphi +\tfrac{m^2}{2} h+m\partial \cdot A=0\label{main2.4}\ , 
\end{align}
\end{subequations}
which can be further simplified combining \eqref{main2.2} and \eqref{main2.4} to reach the following set of equations 
\begin{align}
&\left(\Box-m^2\right)h_{\mu\nu}-2\partial_{(\mu} \partial \cdot h_{\nu)}+\partial_\mu\partial_\nu h=2m\partial_{(\mu}A_{\nu)}+2\partial_\mu \partial_\nu \varphi\ , \label{final1}\\
&\phantom{(}\Box A_\mu -\partial_\mu \partial \cdot A=m\left( \partial_\mu h-\partial \cdot h_\mu \right)\ , \\
&\phantom{(}\Box h-\partial^\mu\partial^\nu h_{\mu\nu}=0\ .\label{final2}
\end{align}
Equations \eqref{final1}--\eqref{final2} correspond to the field equations of linearized massive gravity in the St\"uckelberg formalism, where, as anticipated, $\varphi(x)$ is the St\"uckelberg field and $A_\mu(x)$ is its vector counterpart. \\

A few comments are in order. 
\begin{itemize}
\item
Setting both St\"uckelberg fields to zero, which in literature is known as the \emph{unitary} or \emph{physical gauge}, equations \eqref{main} can be reduced to the Fierz-Pauli system
\begin{align}
\begin{split}
&\left(\Box-m^2\right)h_{\mu\nu}=0\ , \\
&\phantom{(}\partial^\nu h_{\mu\nu} =0\ , \\
&\phantom{(}h=0\ ,
\end{split} 
\end{align}
i.e. the field equations for the theory of a massive spin two field in which the mass term explicitly breaks the gauge invariance.
\item At first sight, the massless $m \to 0$ limit is rather peculiar. Instead of resulting in the free and massless spin two field equations \eqref{free}, it produces the following outcome
\begin{align}
&\Box h_{\mu\nu}-2\partial_{(\mu} \partial \cdot h_{\nu)}+\partial_\mu\partial_\nu h=2\partial_\mu \partial_\nu \varphi \label{E}\ , \\
&\Box A_\mu -\partial_\mu \partial \cdot A=0\label{C}\ , \\
&\Box\varphi =0\label{D}\ .
\end{align}
Equation \eqref{C} represents the field equation for a free-propagating vector field $A_\mu(x)$, describing a spin one massless particle that becomes decoupled in this limit. On the other hand, \eqref{D} describes the propagation of a massless scalar field $\varphi(x)$, which, surprisingly enough, is still coupled to the wanna-be free massless graviton field $h_{\mu\nu}(x)$ in \eqref{E}. This hints at the \emph{vDVZ discontinuity}, a peculiarity associated with the Fierz-Pauli formulation of massive gravity. Notably, there exist other LMG formulations in which this discontinuity is absent, as discussed in \cite{Chamseddine:2018gqh, Gambuti:2021meo, deFreitas:2023ujo} and related literature. The origin of the discontinuity is traced back to the fact that the massless limit of a Fierz-Pauli graviton is not a massless graviton, but rather a massless graviton plus a coupled scalar.\footnote{The resolution lies in the so-called \emph{Vainshtein mechanism}: in a nutshell, general relativity can be recovered around massive bodies by hiding extra degrees of freedom by strong kinetic self-coupling so that they almost do not propagate. We refer to the review \cite{Babichev:2013usa} for further details.} Remarkably, our model captures this characteristic feature of massive gravity.
\end{itemize}

Let us finalize the analysis by investigating the gauge symmetries of the theory. \\

Any theory of massive gravity -- particularly the Fierz-Pauli theory -- is \emph{not} gauge invariant due to the presence of the mass term, unless redundant degrees of freedom are introduced, as is the case for the St\"uckelberg trick. To explore this fact from a first-quantized perspective, it is sufficient to consider the closure equation at ghost number one: while in the massless case, the result is a set of dynamical equations for the diffeomorphism ghost \eqref{dimenticavo}, in the massive scenario things become more intricate due to the presence of another ghost field $\Phi(x)$. The outcome manifests the following additional equations:
\begin{align}
m \, \zeta_\mu = i \partial_\mu \Phi\ , \quad m \, \Phi =0\ .
\end{align}
This indicates that both the scalar and the diffeomorphism ghosts are set to zero, which accounts for the breaking of the gauge invariance due to the presence of a mass term. To display said symmetries, one has to compute the ghost number zero component of $\delta\Psi=\Q \, \Lambda$, where $\Lambda$ is the massive extensions of \eqref{Lambdaparameter}, namely
\begin{equation} \label{Lambda}
\Lambda=i\varepsilon_\mu(x)\,(\psi^\mu_1\beta_2-\psi^\mu_2\beta_1)+\Sigma(x)\,(\theta_1\beta_2-\theta_2\beta_1)+\cdots\ ,
\end{equation}
where $\varepsilon(x)$ and $\Sigma(x)$ are the two gauge parameters associated with the two gauge symmetries. The outcome is
\begin{align}
& \delta h_{\mu\nu}=2\partial_{(\mu} \varepsilon_{\nu)} \quad \delta A_\mu =-m \, \varepsilon_\mu\ , \\
& \delta A_\mu =\partial_\mu \Sigma \qquad \;\; \delta \varphi =-m \, \Sigma\ ,
\end{align}
as expected from the St\"uckelberg formulation. Therefore, the massive $\N=4$ spinning particle produces a gauge invariant formulation of LMG.

\subsubsection{Auxiliary oscillators approach}
The main idea of the auxiliary oscillators approach consists of enlarging the BRST algebra with additional variables, whose physical interpretation can be seen as to describe the internal degrees of freedom of the $\N=4$ spinning particle. Thus, the first step involves an extension of the phase space by introducing two canonically conjugated complex bosonic variables, denoted as $(\alpha^a, \bar{\alpha}_a)$, with commutation relation 
\begin{equation}
[\bar{\alpha}_a,\alpha^b] = \delta^b_a\ .
\end{equation}
Note that the index $a$ may run over an arbitrary set, including a single value. To preserve the $\mathcal{N}=4$ worldline supersymmetry, the introduction of four complex fermionic variables $(\eta_i^a,\bar{\eta}^i_a)$ is necessary. They represent the superpartners of the $\alpha$'s, with anticommutator
\begin{equation}
\{\bar{\eta}^i_a , \eta_j^b \} =\delta^i_j \delta^b_a\ .
\end{equation}
The next step involves finding a way to deform the SUSY constraints while ensuring the closure of the algebra. A straightforward attempt would be to extend the $\N=2$ constraints of \cite{Carosi:2021wbi}, resulting in
\begin{align}
\begin{split}
H&=\tfrac12 \left(p^2+ m^2 \, \alpha^a\bar{\alpha}_a + m^2 \, \eta^a \cdot \bar{\eta}_a\right)\ , \\
q_i&=p \cdot \psi_i + m \, \eta_i^a\bar{\alpha}_a\ , \\
\bar q^i&=p \cdot \bar \psi^i + m \,\bar{\eta}^i_a \alpha^a\ .
\end{split}
\end{align}
However, this leads to $\left\{q_i,\bar q^j \right\} \neq 2\delta_i^j \, H$, specifically
\begin{equation} \label{prob}
\left\{q_i,\bar q^j \right\}=\delta_i^j p^2+m^2\left( \eta_i^a \bar \eta^j_a +\delta_i^j\alpha_a\bar\alpha^a\right)
\end{equation}
and there is not an obvious redefinition of the Hamiltonian that allows for a closure. Consequently, the resulting algebra is not first-class, and potential issues are anticipated in the quantization process. Indeed, the associated BRST charge $\Q= c\,H+\gamma \cdot \bar q+\bar\gamma \cdot q-2\bar\gamma \cdot \gamma\, b$ fails to be nilpotent:
\begin{equation} \label{ivo}
\Q^2=m^2 \left( \bar \gamma \cdot \eta^a \gamma \cdot \bar \eta_a -\gamma \cdot \bar \gamma \, \eta^a \cdot \bar \eta_a \right)\neq 0\ .
\end{equation}
The auxiliary oscillators approach may be a viable option, although it is necessary to follow the principles outlined in section \ref{sec3.2}. One should engineer a deformation of the BRST algebra that leads to preventing the obstruction given by \eqref{ivo}, e.g. through a suitable redefinition of the Hamiltonian and of the $so(4)$ constraints. Further work is needed to explore this avenue, but such investigations are beyond the scope of the current discussion, as a promising method for dealing with a massive graviton is already at hand.


\section{Massive graviton on curved spacetimes} \label{sec4}
In this section, the investigation of a first-quantized massive graviton on a curved spacetime is carried out. Previous works have addressed the challenge of coupling massive higher spin fields -- which, in the present formalism means for $\N>2$ -- but the analysis failed in going beyond (A)dS spaces \cite{Bastianelli:2014lia}. The goal of the present work is to overcome said result with the $\N=4$ spinning particle. Let us rephrase the objective more clearly: the aim is to provide a worldline formulation of the \emph{linear} theory of massive gravity, namely of the Fierz-Pauli theory, on a curved spacetime; such a theory describes the propagation of massive spin $2$ particle on a non-flat background. \\
The coupling to a generic curved background is achieved through the covariantization of the reduced model previously discussed, which results only in a mild modification of the massless model of section \eqref{sec3.2}. The supercharges are deformed as
\begin{equation}
\q_i:=-i\,\psi_i^a\,e^\mu_a\,\hat \nabla_\mu\ +m\theta_i\ ,\quad \bar \q^i:=-i\,\bar\psi^{i\,a}\,e^\mu_a\,\hat \nabla_\mu+m\bar\theta^i\ ,
\end{equation}
and together with the Laplacian $\nabla^2$ form the same obstructed algebra \eqref{bo}, with the sole modification arising from the following anticommutator
\begin{equation}
\{\q_i, \bar \q^j\}=-\delta_i^j\left(\nabla^2-m^2\right)-\psi^\mu_i\bar\psi^{\nu\,j}\, \bm{R}_{\mu\nu}\ .
\end{equation}
The deformed BRST charge displays still the general form $\Q=c\,\D+\bm{\nabla}+\bar\gamma \cdot\gamma\,b$, where now
\begin{align} 
\D:=\nabla^2-m^2+\Re\ , \quad &\text{with} \quad \Re:=R_{\mu\nu\lambda\sigma}\,\psi^\mu \cdot\bar\psi^\nu\,\psi^\lambda \cdot\bar\psi^\sigma+ \tfrac{1}{2}R\ , \label{DD} \\
\bm{\nabla}:=-i\,S^\mu\hat \nabla_\mu+ m\rho\ , \quad &\text{with} \quad \rho:=\bar \gamma \cdot \theta+\gamma \cdot \bar \theta\ .
\end{align} 
The model is considered on Einstein spaces: indeed, it is expected for the massive theory to reproduce the correct results in the $m \to 0$ limit -- similarly to the flat spacetime scenario -- and the massless BRST charge is nilpotent only when $R_{\mu\nu}=\tfrac14 g_{\mu\nu}R$. It is not possible to obtain a weaker condition for the background in the massive case. Moreover, from a QFT perspective, it is known that an Einstein space is the only space on which a free (massive) graviton can consistently propagate \cite{Aragone:1971kh, Bengtsson:1994vn, Buchbinder:1999ar, Deser:2006sq}. Hence, it is natural to implement this condition right at the beginning of the analysis. \\

The $so(4)$ constraints ${\cal T}_\alpha$ are given in \eqref{Ta}, thus the wavefunction $\Psi$ can be read from \eqref{BVstringfield2}. This section aims to investigate if and how the presence of the mass affects the nilpotency of the BRST operator. In contrast to the case of the massive spin one particle on a general background, where nilpotency was not a concern even in the massless case, the situation is different here. The massless $\N=4$ spinning particle already selects only a specific subset of available backgrounds. The starting point remains
\begin{equation}
\Q^2=\bm{\nabla}^2+\bar\gamma \cdot\gamma\,\D+c\,\left[\D, \bm{\nabla}\right]\ , 
\end{equation}
and thus, it is necessary to address the independent obstructions
\begin{align} 
\bm{\nabla}^2+\bar\gamma \cdot\gamma\,\D &= -\tfrac12\,S^\mu S^\nu\, \bm{R}_{\mu\nu}+m^2\left(\cancel{ \gamma \cdot \bar \theta \, \gamma \cdot \bar \theta + \bar \gamma \cdot \theta \, \bar \gamma \cdot \theta }\right)+\bar\gamma \cdot\gamma\,\Re\ , \label{first obstruction22} \\ 
\left[\D, \bm{\nabla}\right] &= -iS^\mu\nabla^\lambda \bm{R}_{\lambda \mu} +i S^\mu\nabla_\mu \bm{R}+\kappa [\,R,\bm{\nabla}]\ . \label{second obstruction23}
\end{align}
Just as in the case of the Proca theory, it is evident that the mass does not enter into the BRST algebra. Specifically, note that the term in \eqref{first obstruction22} is canceled both from a symmetry-based argument and by the fact that it annihilates any states $\Psi$ in the physical sector, thereby being set to zero. Additionally, no mass term survives inside the commutator in \eqref{second obstruction23}. Consequently, the mass does not \emph{explicitly} obstruct the nilpotency. The expressions remain the same as in the massless case: once evaluated on the reduced Hilbert space $\mathcal{H}_{\mathrm{red}}$, taking Einstein manifold simplifications into account, they read
\begin{align}
\bm{\nabla}^2+\bar\gamma \cdot\gamma\,\D &\stackrel{{\ker}{\cal T}_\alpha}{=}-\frac12\,\bar\gamma\cdot\gamma\left(\psi^\mu \cdot\bar\psi_\mu-1 \right)R\ , \label{first-first2} \\
\left[\D, \bm{\nabla}\right] &\stackrel{\ker{\cal T}_\alpha}{=}0\ . \label{second-second2}
\end{align}
The leftover operator \eqref{first-first2} is vanishing when acting on the massless physical wavefunction $\Q^2\left. \Psi\right|_{m=0}=0$. However, this is \emph{not} the case for the massive graviton, as the wavefunction has changed. When acting on the physical sector, the outcome is
\begin{align}
\Q^2\Psi(x,\psi,\theta\,|\,\g)=-\tfrac{1}{2} \, \phi \left( \theta_1\gamma_2-\theta_2\gamma_1 \right)cR-\tfrac{1}{2} \, \Phi \left( \theta_1\gamma_2-\theta_2\gamma_1 \right)R\neq 0\ .
\end{align}
Let us highlight that attempts to implement possible couplings involving traces of the Riemann tensor and the newly introduced fermionic variables inside the Hamiltonian \eqref{DD}, such as
\begin{equation}
\Delta R=c_1 \, R \, \theta \cdot \bar \theta + c_2 \, R \, \theta \cdot \bar \theta \, \theta \cdot \bar \theta+ c_3 \, R_{\mu\nu} \, \psi^\mu \cdot \bar \psi^\nu \theta \cdot \bar \theta\ , 
\end{equation}
do not solve the issue. Although an appropriate tuning of the coefficients -- in particular one finds $c_1+c_2=-2$ -- potentially addresses the first obstruction \eqref{first-first2}, these terms would disrupt the second one \eqref{second-second2}, compelling $c_1=c_2=c_3=0$. This explains why those terms have been omitted right from the beginning: as already discussed in \cite{Bonezzi:2018box}, one cannot have fermions inside $\Delta R$. The only possibility to achieve nilpotency of the BRST operator is to consider Ricci-flat backgrounds, i.e.
\begin{equation}
R_{\mu\nu}(x)=0\ ,
\end{equation}
which immediately addresses the remaining obstruction \eqref{first-first2}, yielding a nilpotent BRST charge. This outcome resembles the scenario encountered in string theory, where consistency of the string propagation, worldsheet conformal invariance, implies Ricci flatness to leading order \cite{Lu:2003gp}. Such an unfortunate result amounts to the difficulties that arise when trying to give a mass to a graviton and trying to generalize it to a curved spacetime. As discussed in the introduction, several obstacles have been encountered on the quantum field theory side, with a promising solution emerging only in recent years. It is plausible that a similar intricate path has to be navigated even within the worldline formalism. The present work may then serve as an initial step in the direction of future developments. \\

In conclusion, the field equations for a massive graviton on an Einstein background with zero cosmological constant are evaluated using the deformed BRST operator incorporating the appropriate curvature terms. Upon eliminating auxiliary fields, the resulting equations of motion are
\begin{subequations} \label{aaa}
\begin{align}
&\left( \nabla^2-m^2 \right)h_{\mu\nu}-2\nabla_{(\mu}\nabla \cdot h_{\nu)}+\nabla_\mu\nabla_\nu h+2R_{\mu\alpha\nu\beta}h^{\alpha\beta}=2m \nabla_{(\mu}A_{\nu)}+2\nabla_\mu\nabla_\nu\varphi\ , \\
&\left( \nabla^2-m^2 \right)h-\nabla^\mu\nabla^\nu h_{\mu\nu}=2m \nabla \cdot A+2\nabla^2\varphi\ , \\
&\phantom{(}\nabla^2 A_\mu-\nabla_\mu\nabla\cdot A=m(\nabla_\mu h -\nabla \cdot h_\mu)\ , \\
&\phantom{(}\nabla^2 \varphi +\tfrac{m^2}{2}h+m \nabla \cdot A=0\ ,
\end{align}
\end{subequations}
which can be simplified as
\begin{align}
&\left( \nabla^2-m^2 \right)h_{\mu\nu}-2\nabla_{(\mu}\nabla \cdot h_{\nu)}+\nabla_\mu\nabla_\nu h+2R_{\mu\alpha\nu\beta}h^{\alpha\beta}=2m \nabla_{(\mu}A_{\nu)}+2\nabla_\mu\nabla_\nu\varphi\ , \label{final2.1} \\
&\phantom{(}\nabla^2 A_\mu-\nabla_\mu\nabla\cdot A=m(\nabla_\mu h -\nabla \cdot h_\mu)\ , \\
&\phantom{(}\nabla^2 h-\nabla^\mu\nabla^\nu h_{\mu\nu}=0\ .\label{final2.2}
\end{align}
Equations \eqref{final2.1}--\eqref{final2.2} correspond to the field equations governing linearized massive gravity on a Ricci-flat background within its St\"uckelberg formulation. The same equations of motion can be derived from a field theory approach: one has to take into account the most general action functional while keeping only a restricted class of non-minimal couplings to the background, namely, the ones that do not excite possible unphysical degrees of freedom \cite{Buchbinder:1999ar}. The final step involves performing the necessary St\"uckelberg tricks to restore gauge invariance. As a final remark, it is worth noting that the correct gauge symmetries can be derived from $\Lambda$ \eqref{Lambda}, yielding
\begin{align}
& \delta h_{\mu\nu}=2\nabla_{(\mu} \varepsilon_{\nu)} \quad \delta A_\mu =-m \, \varepsilon_\mu\ , \\
& \delta A_\mu =\nabla_\mu \Sigma \qquad \; \; \delta \varphi =-m \, \Sigma
\ .
\end{align}


\section{Conclusions} \label{conc}
In this work, we have investigated the issue of consistently coupling massive higher-spin fields to a general curved background within the wordline formalism, with our primary focus being that of a first-quantized formulation of linearized massive gravity. 
During this investigation, we have explicitly shown the BRST quantization of the massive $\N=2$ spinning particle coupled to off-shell gravity, which, to the best of our knowledge, has never been investigated before in the literature with these tools. 
Our results show that the model reproduces the Proca theory on curved spacetime with the specific selection of the minimal coupling to the background, improving knowledge built on previous results and offering potential utility in the context of worldline computations employing this model. Subsequently, we have addressed directly the issue of a first-quantized massive graviton, discussing the correct reproduction of the Fierz-Pauli theory on a flat spacetime through the dimensional reduction of the higher-dimensional massless $\N=4$ spinning particle. Additionally, we addressed the challenges associated with the auxiliary oscillators approach. Finally, we tried to couple the massive particle to an Einstein background. Our findings suggest that a Ricci-flat spacetime emerges as the only available background for consistency at the quantum level. Let us stress that the theory constructed here is the worldline counterpart of the \emph{linear} theory of massive gravity, namely the Fierz-Pauli theory from a QFT perspective: section \ref{sec4} has been devoted to the correct reproduction, starting from the BRST system of the spinning particle model \eqref{BRSTsystem}, of the field equations and the gauge symmetries of LMG in the St\"uckelberg formulation. In particular, from the quantum field theory side, the same results can be derived from the FP action put on a curved spacetime, which is the theory for a massive spin $2$ propagating on non-flat geometry and which can be obtained as the first approximation of a general massive gravity theory around a given fixed background. Even from the quantum field theory side, the construction of interacting theories of massive gravity is much harder, and a well-defined non-linear completion of the theory has been only recently constructed: achieving such a result in the first quantization formalism -- if feasible -- would require a considerable effort and is beyond the scope of the present work. \\
With these results at hand, this work may be regarded as a first step towards the realization of a first-quantized massive graviton on an Einstein spacetime with \emph{non-zero} cosmological constant, either by identifying consistent improvements in the dimensional-reduced BRST system, or fully pursuing the auxiliary oscillators procedure, contingent on resolving the seemingly simpler yet already intricate flat spacetime scenario first. As a preliminary step in both instances, an intriguing prospect involves investigating whether modifying the model could result in a first-quantized \emph{partially massless} graviton \cite{Higuchi:1986py, Deser:2001us}. There are possible extensions worth exploring, such as relaxing the constraint on the BRST Hilbert space to subalgebras, which should in principle lead to the inclusion inside the BV spectrum of the $\N=0$ supergravity, i.e. the particle theory that has in its spectrum the graviton, the dilaton, and the antisymmetric Kalb-Ramond tensor field. Such extensions might be pursued along the lines of \cite{Bonezzi:2020jjq}, investigating possible couplings to the background fields through deformations of the BRST charge. 
Furthermore, the massive $\N=4$ worldline model developed here could be exploited in future computations. One potential application involves investigating the one-loop divergences of massive gravity, reproducing, and potentially extending, the results recently obtained from the quantum field theory side \cite{Dilkes:2001av, Ferrero:2023xsf}. To pursue this application, it is necessary to establish a suitable one–loop quantization of the massive model and study the corresponding partition function on the one–dimensional torus. The initial step would be to determine the correct measure on the moduli space of the supergravity multiple, implementing the correct projection on the massive gravity contribution, following the analysis of \cite{Bastianelli:2019xhi}. The computation of the heat kernel/effective action coefficients should proceed along the lines of \cite{Bastianelli:2023oca}, where the counterterms necessary for the renormalization of the one-loop effective action of pure gravity were computed using the massless $\N=4$ spinning particle. This project is currently underway \cite{Fecit:2024}.

\acknowledgments
I am deeply grateful to R.~Bonezzi and I.~Sachs for helpful discussions and insightful suggestions at various stages of this research. Additionally, I am thankful to F.~Bastianelli for providing valuable input and offering constructive feedback on the draft. I further thank V.~Larotonda for careful draft reading. I would like to express gratitude to the organizers and the students of the 29th W.E. Heraeus Summer School ``Saalburg" for Graduate Students on ``Foundations and New Methods in Theoretical Physics" for generating an inspiring scientific environment, which led to meaningful discussions. I also acknowledge hospitality during the final stage of the work by the LMU Munich.


\bibliographystyle{jhep}
\bibliography{biblio.bib}


\end{document}